\documentstyle[prb,aps,epsf,twocolumn]{revtex}
\begin{document}
\title{\large Mesoscopic Charge Density Wave in a Magnetic Flux   }
\author{Gilles Montambaux}
\address{Laboratoire de Physique des Solides,  associ\'e au
CNRS \\ Universit\'{e} Paris--Sud \\ 91405 Orsay, France}
\twocolumn[
\date{\today}
\maketitle
\widetext
 \begin{center}
 \begin{abstract}
\parbox{14cm}{The stability of a Charge Density Wave (CDW) in a
 one-dimensional ring pierced by a Aharonov-Bohm flux is studied in a
mean-field picture. It is found that the stability depends on the parity of
the number $N$ of electrons. When the size of the ring becomes as small as
the coherence length $\xi$, the CDW gap increases for even $N$ and decreases
for odd $N$. Then when $N$ is even, the CDW gap decreases with flux but it
increases when $N$ is odd.
The variation of the BCS ratio with size and flux is also calculated.
We derive the harmonics expansion of the persistent current in
a presence of a finite gap.
} \end{abstract}
\end{center}
%\pacs{PACS Numbers: 73.40 H, 72.15 G, 72.15 N}
]
\narrowtext
\section{Introduction}\label{INT}

It has been proposed recently that a  Aharonov-Bohm (AB) flux $\phi$
should affect the stability of a Charge
Density Wave (CDW) in a one-dimensional  ring
  geometry \cite{Bogachek90,Visscher96,Yi97} as it may have been seen  in a
recent experiment on CDW pierced by AB flux lines trapped in columnar
defects\cite{Latyshev97}.
One possible cause for the modulation of the CDW stability is
 the discreteness of the spectrum
and  the modulation of the position of the energy levels with the
flux\cite{Visscher96,Yi97}. Indeed, it has been predicted that the CDW
gap and the critical temperature
 oscillate with the flux $\phi$, with a period $\phi_0=h/e$ and that they
are maxima at
$\phi=0$ and  minima at $\phi=\phi_0/2$. When the perimeter $L$ of the ring
 becomes of the order of the correlation length $\xi$, i.e. when the CDW
order parameter becomes of the order of the mean level spacing, the CDW
 can even be destroyed when $|\phi - n\phi_0|$  is larger than a
critical value\cite{Visscher96,Yi97}.

In this paper, we elaborate on these ideas and we show that the stability of
the
CDW depends  crucially on the parity of the number $N$ of particles in the
1D ring
(considering here spinless particles), an effect which has not been
properly
considered in previous works. When this number $N$ is odd, the CDW
is destabilized by decreasing the length of the ring. But when $N$ is even,
it we find that the CDW is stabilized.
When $N$ is even,  the effect of an AB flux is to destabilize the
CDW, as found in   refs.\cite{Visscher96}. But when  $N$ is {\it odd},
 we find that the
CDW can be {\it stabilized by the flux},
 contrary to the conclusions of refs.\cite{Visscher96,Yi97}.
This effect is reminiscent of the parity
effect for the
 persistent current in one-dimensional
rings\cite{Trivedi88,Cheung89,Leggett91}.
In the following, we shall consider the case of spinless electrons.

   In the next section, we establish the thermodynamic equations
for the CDW in a finite 1D system, with emphasis on the parity
effect. In section III, we calculate the flux variation of the order
parameter in the form of a Fourier series. In section IV, the same is done
for the critical temperature. Finally we calculate the persistent
 current in the last section.

\section{CDW in a magnetic flux} \label{CDW}

\begin{figure}[hbt]
\centerline{
\epsfxsize 6cm
\epsffile{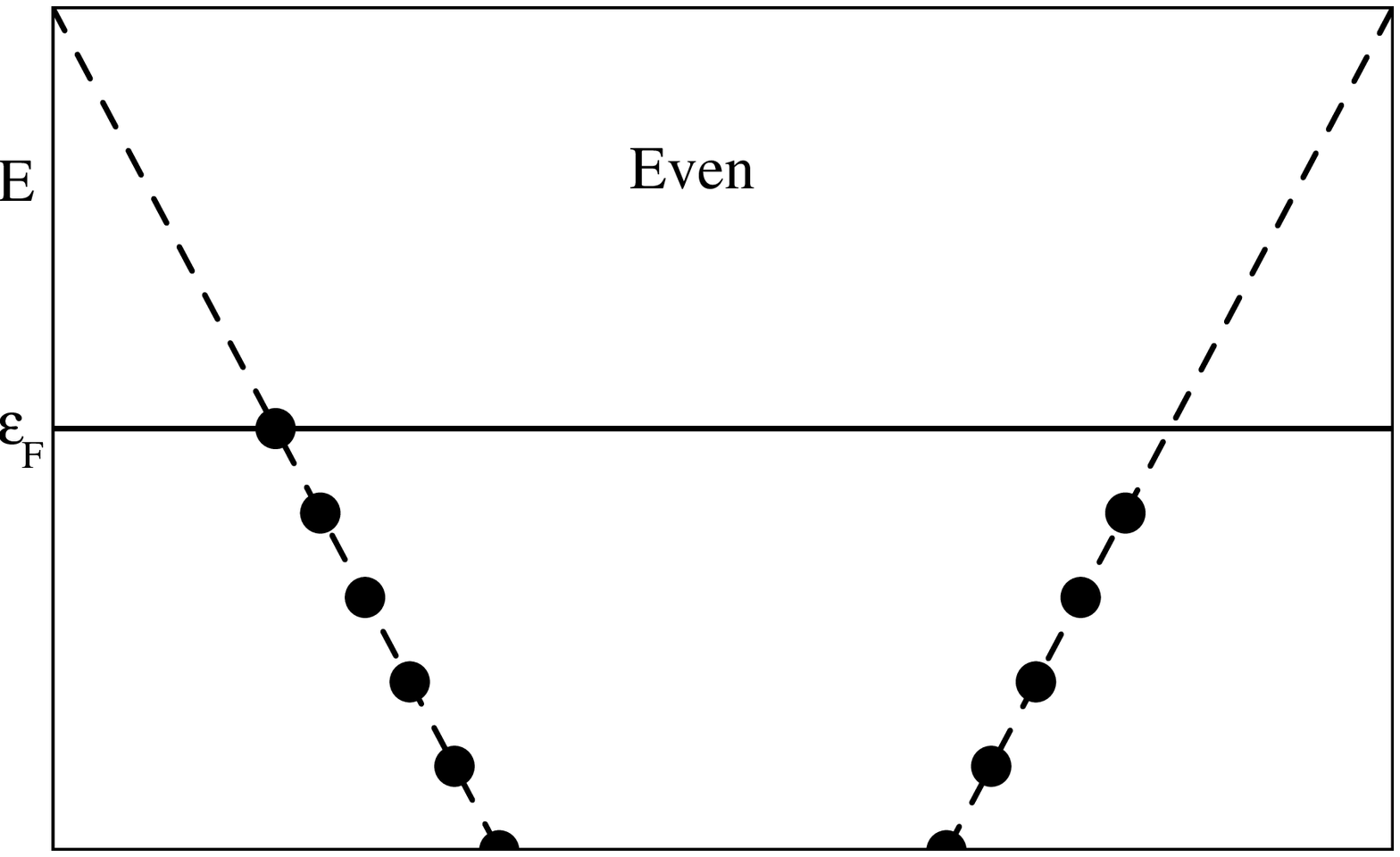}}
\centerline{
\epsfxsize 6cm
\epsffile{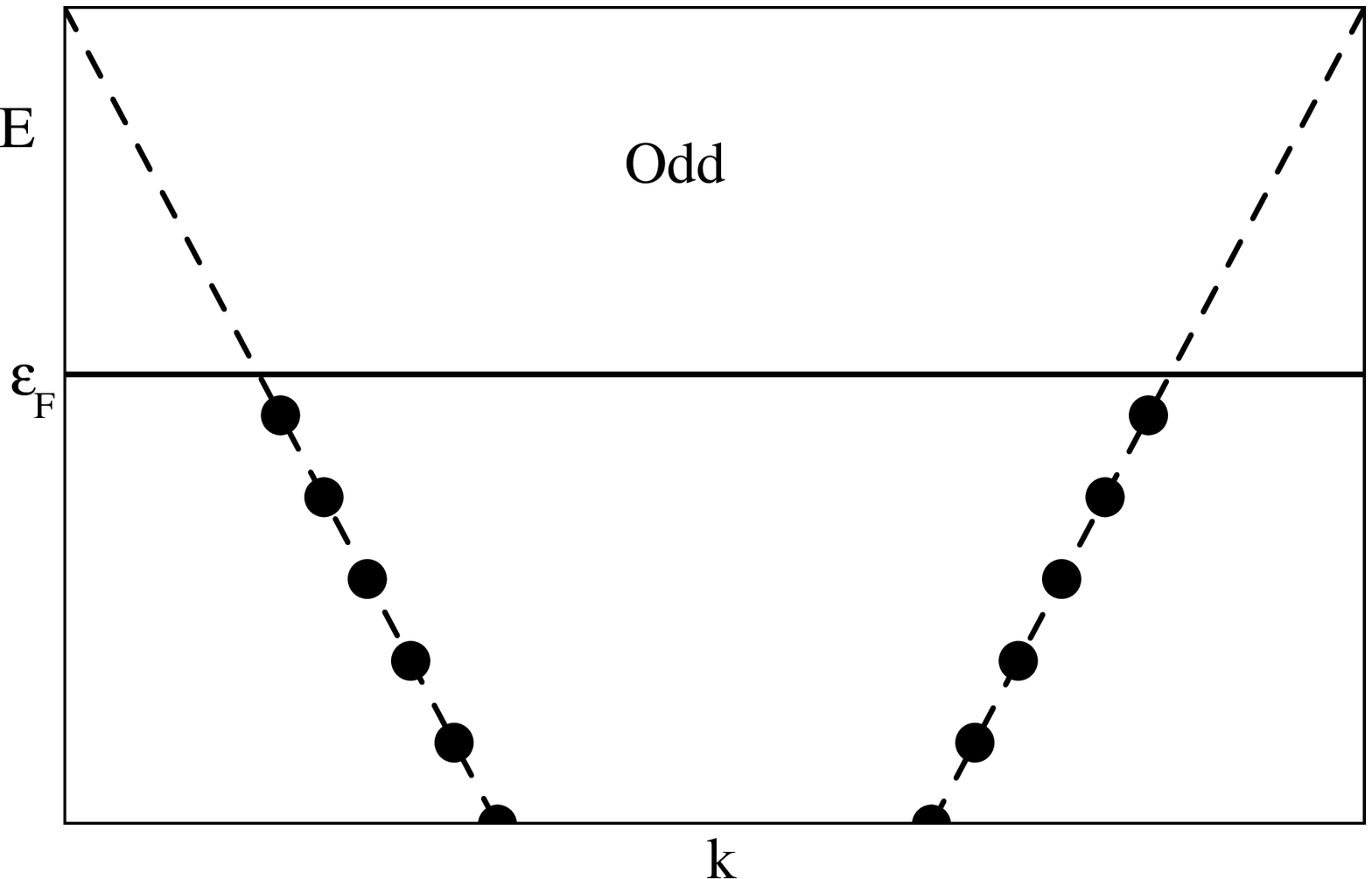}}
\caption{Dispersion relation in the metallic phase, a) when the
number $N$
of particles is even (do not forget the state $k=0$), b) when $N$ is odd.
The horizontal line gives the position of the Fermi level.}
\label{fig1}
\end{figure}

Consider a one-dimensional ring of perimeter $L$.
The periodic boundary conditions fix the wave vector of
 the eigenstates. In the presence of a flux, the possible wave vectors  are:$$k={2 \pi \over L} (p+\varphi)$$
where $\varphi$ is the dimensionless flux $\phi/\phi_0$ and $p
\in {\cal{Z}}$.
Let $N$ be the total number of electrons. The Fermi wave vector $k_F$ is:
$$k_F ={N \pi \over L}$$ and the Fermi energy $\epsilon_F$ is:
\begin{equation}\epsilon_F= {\hbar^2 \over 2 m} \left({N \pi \over L}\right)^2\ \ \ .\label{EF}\end{equation}
It is important to
stress that with this choice of $k_F$ and $\epsilon_F$, {\it the number of
electrons is independent of the flux}, Fig.(\ref{fig1}).

Although the chemical potential is in principle fixed by a reservoir in
ref. \cite{Visscher96}, we shall see later that their results
 correspond actually to a fixed {\it even} number of particles.
 \medskip

\begin{figure}[hbt]
\centerline{\epsfxsize 6cm
\epsffile{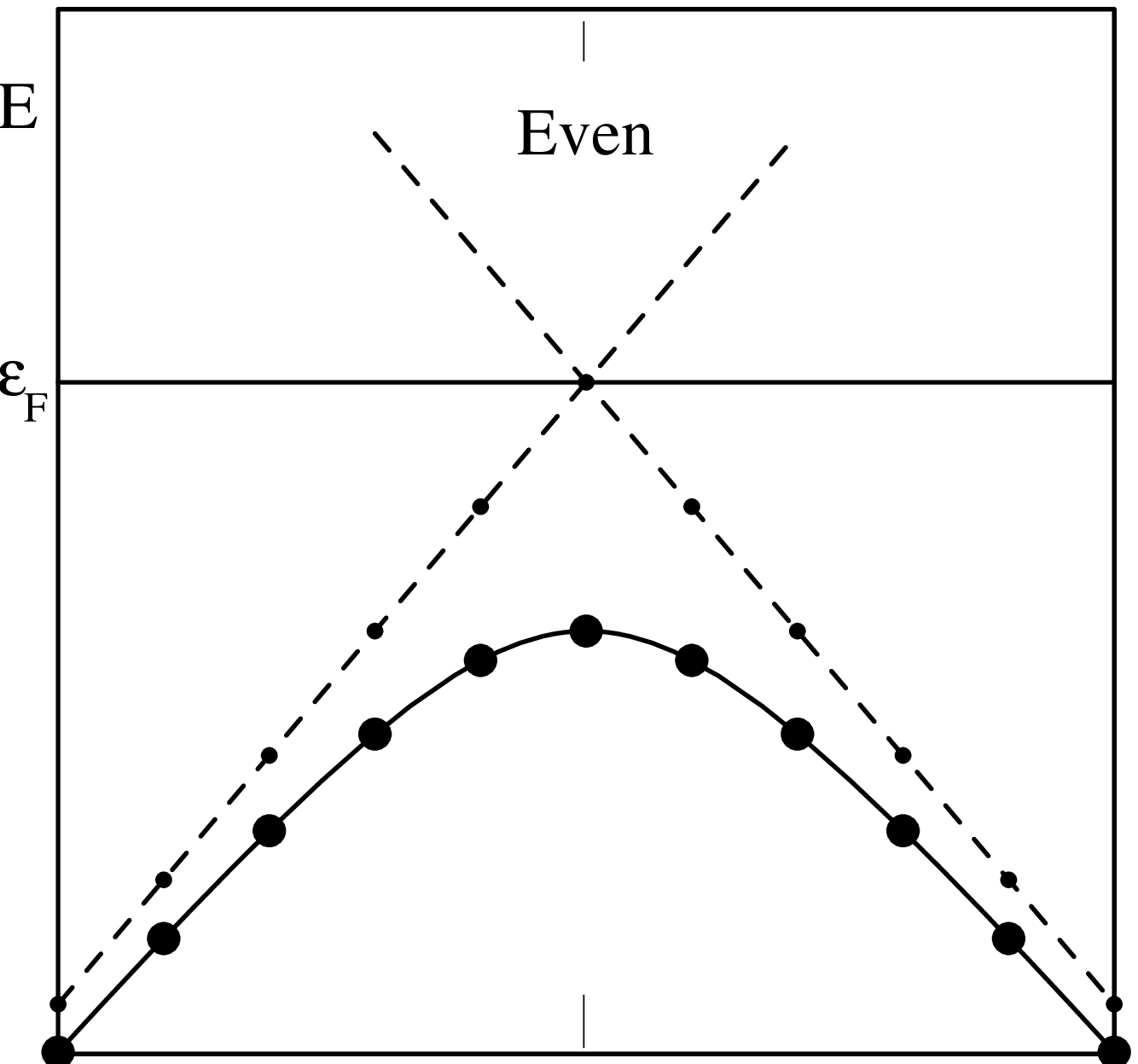}}
\centerline{\epsfxsize 6cm
\epsffile{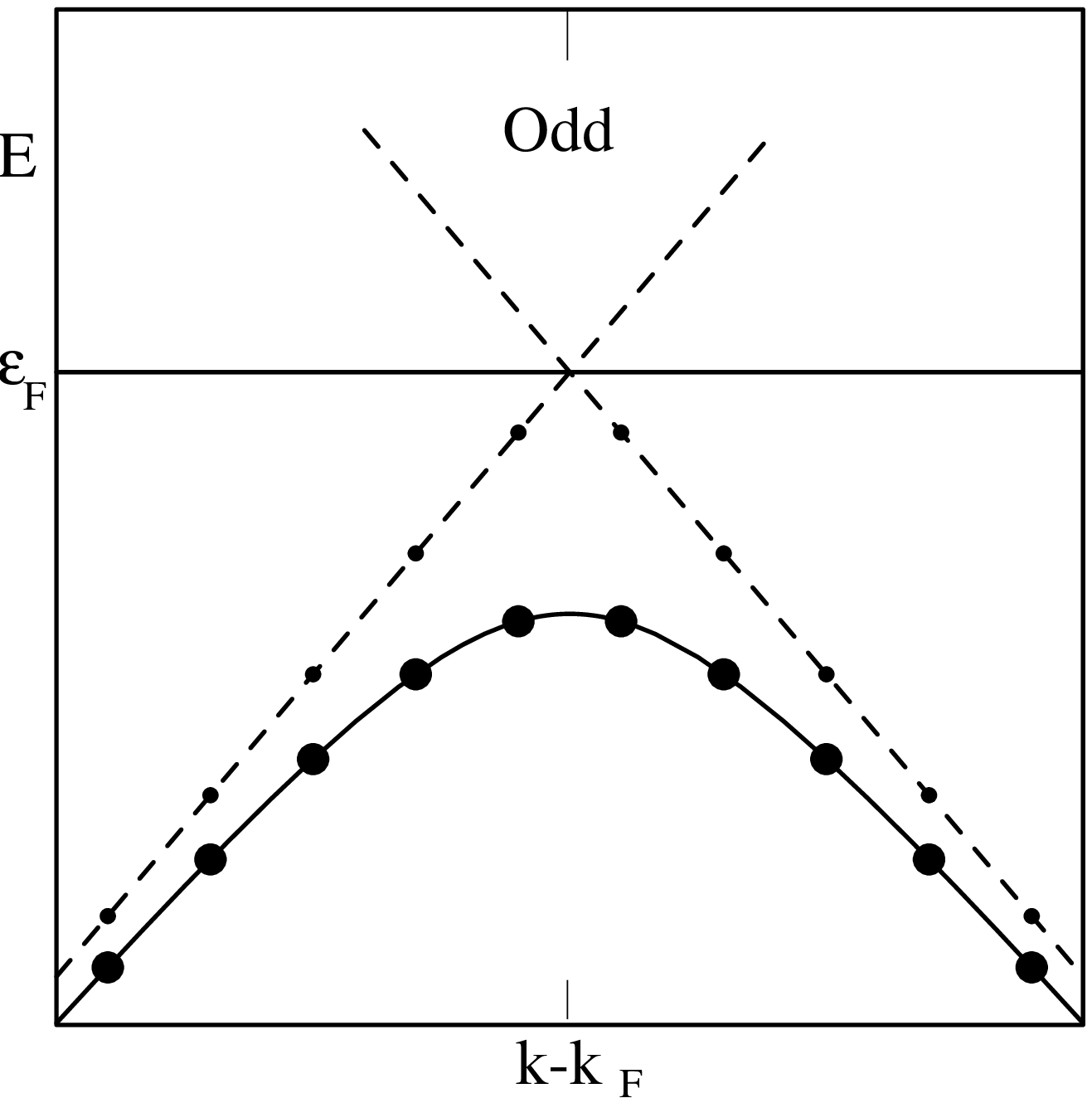}}
\caption{Dispersion relation near the Fermi level, a) when
the number $N$
of particles is even , b) when $N$ is odd. The
horizontal line gives the position of the Fermi level. The small dots
denote the states in the metallic phase and the large dots represent
the states in the CDW phase}
\label{fig2}
\end{figure}

The theory of the CDW ordering is well known. In the presence of a periodic
potential with wave vector $Q$, the Hamiltonian is:
$${\cal{H}}= {\cal{H}}_0 + 2 \Delta \cos(Q r +\theta)$$
where the eigenenergies $\epsilon_k$ of  ${\cal{H}}_0$ are known.
Because of the periodic potential, the states $k$ are in principle coupled
to states $k\pm n Q$ where $n=1,2,3,\dots$.
We consider the case of a  weak potential. Neglecting commensurability
effects,
 one has to diagonalize a $2 \times 2$ matrix: the states with wave
vector $k \simeq k_F$ are coupled to states with wave vector $k-Q \simeq
- k_F$: $$\left( \begin{array}{lc}
\epsilon(k) &      \Delta e^{i \theta}  \\
 \Delta e^{-i \theta}  &  \epsilon(k-Q) \end{array} \right)$$
The eigenvalues are given by $E={\epsilon_k+\epsilon_{k-Q} \over 2} \pm
\sqrt{\left({\epsilon_k-\epsilon_{k-Q} \over 2}\right) ^2 + \Delta^2}$.
It is then convenient to linearize the dispersion near the Fermi level so that, by taking the origin of the energies at the Fermi level, one has:
$$\epsilon_k=\hbar v_F (|k|-k_F) = 2 \delta \ \left(|p+\varphi|-{N \over
2}\right)$$ where the mean level spacing $\delta$ near the Fermi level is
given by $\pi \hbar v_F
/ L$. Writing the nesting vector $Q$ in the form $Q=2 \pi q /L$, where $q$ is
an integer, the eigenvalues in the CDW phase are:
$$E_p=  (q -N)\delta   \pm\sqrt{4 \delta^2 (p-{q \over
2}+\varphi)^2+\Delta^2}\ \ \ .$$  The nesting condition implies $q=N$ so
that the nesting vector is  $$Q=2 k_F={2 N \pi \over L}$$
and  the energy levels in the CDW phase become:
\begin{eqnarray}E_p&=& \pm \sqrt{4 \delta^2 (p-{N \over
2}+\varphi)^2+\Delta^2} \nonumber \\ &=&
\pm 2 \delta \sqrt{(p-{N \over 2} + \varphi)^2 +\Lambda^2}\end{eqnarray}
 where $\Lambda= \Delta / (2\delta)$. $\Lambda$ measures the gap amplitude
in units of the mean level spacing  and can be written
as the ratio of the length over the CDW coherence length $\xi=\hbar v_F / \pi
\Delta$. One has $\Lambda = L/ (2 \pi^2 \xi)$.
The effects discussed in this paper appear for small rings when the
mean level spacing becomes as large as the the CDW gap, see
Fig.(\ref{fig2}). 

From minimization of the
total free energy $$ F = {\Delta^2 \over \lambda} L - k_B T Log {\cal{Z}}$$
where $\lambda$ is the interaction parameter and $\cal{Z}$ is the grand
canonical partition function,  the well-known self-consistency condition is
obtained: $$1  =  \sum_p {g \over 2 E_p}\tanh ({\beta \over 2} E_p)
$$
where $g = \lambda / 2 \pi \hbar v_F$.
When $N$ is even, the introduction of a new variable $n=p-N/2$ reduces this
equation to:
\begin{equation}1 = g  \sum_n {\tanh ({\beta' \over 2}
\sqrt{(n+\varphi)^2+\Lambda^2}) \over 2 \sqrt{(n+\varphi)^2+\Lambda^2} }
\label{gapeven}\end{equation}
where $\beta'=2 \beta  \delta$. This is the result  found in
ref.\cite{Visscher96}. However, when the number $N$ is odd, $n=p-(N-1)/2$
and the self-consistency equation becomes:
\begin{equation}1 = g \sum_n {\tanh ({\beta' \over 2}
\sqrt{(n-{1 \over 2}+\varphi)^2+\Lambda^2}) \over
2 \sqrt{(n-{1 \over 2}+\varphi)^2+\Lambda^2} }\label{gapodd}\end{equation}
 Thus {\it the  stability of the CDW depends on
the parity of the number of electrons}. The flux
dependence for the even and odd parities are deduced from each
other by a
translation of $\phi_0/2$. We stress the fact that the results of
ref.\cite{Visscher96} correspond to a fixed {\it even} number of particles,
once
the chemical potential is fixed to the value of eq.(\ref{EF}). If the
chemical potential was fixed to any other value, the number of particles would vary with the field leading to discontinuities of the persistent current \cite{Cheung89} and of the CDW
 stability.
\section{Ground state}  \label{GRO}

We first study the evolution of the
CDW order parameter $\Delta_\varphi$
 at zero temperature, as a function of the size and of the AB flux. When
$N$ is even, it is given by
\begin{equation}
1 = {g \over 2} \sum_n {
1 \over  \sqrt{(n+\varphi)^2+\Lambda_\varphi^2} }\label{gap1}
\end{equation}
where the dimensionless parameter
$\Lambda_\varphi=\Delta_\varphi / 2 \delta$ has been introduced. This sum
diverges at large $n$ and is usually cut-off at an energy scale
$E^*$ of the order of the bandwidth.  This corresponds to an integer
$n^*=E^*/2\delta$. In ref.\cite{Visscher96}, the flux dependence of the
order parameter is given by:

\begin{equation} \sum_n[ {
1 \over  \sqrt{(n+\varphi)^2+\Lambda_\varphi^2} }- {
1 \over  \sqrt{n^2+\Lambda_0^2} }] =0\label{gap2}\end{equation}
which corresponds to the case where $N$ is even. However, when $N$ is odd,
 $n$ should
be replaced by $n-1/2$ as seen in eq.(\ref{gapodd}).
\medskip

 We found it
convenient to
write the Fourier decomposition of the flux
 dependence of the order parameter. To transform eq.(\ref{gap1}), we use the
Poisson summation formula $$\sum_n f(n+\varphi) =g_0+ 2\sum_{m>0} g_m \cos
(2 \pi m \varphi)$$ where$$g_m=2\int_{0}^{\infty} f(y) \cos (2 \pi m y) dy\ \ \ .$$
The constant term $g_0$ is divergent and must be cut-off with  $n^*$.
 The Fourier components are convergent. As a result one gets:
\begin{equation} \label{gap3}{1 \over g}= \ln {2 E^* \over \Delta_\varphi}
+ 2 \sum_{m>0} (-1)^{Nm} K_0(2 \pi m \Lambda_\varphi) \cos 2 \pi m \varphi
\end{equation} The factor $(-1)^{Nm}$ in the harmonics expansion takes into
account the parity effect. $K_0$ is a modified Bessel function of the second
kind\cite{Bessel}. \bigskip

It is instructive to consider first the ring in the absence of
external flux.
For the infinite bulk system, the gap $\Delta_b$  is given by:
\begin{equation} \label{gap4}{1 \over g}= \ln {2 E^* \over
\Delta_b}\end{equation}
When the size becomes finite, the gap in zero flux $\Delta_0$ is given by:
\begin{equation}
\ln {\Delta_0 \over \Delta_b} = 2 \sum_{m>0}
(-1)^{Nm} K_0(2 \pi m {\Delta_0 \over \Delta_b} \Lambda_b)
\label{delta0}
\end{equation}
where the parameter $\Lambda_b=\Delta_b /2 \delta$ measures the bulk gap in
units of mean level spacing and is proportional to the size.

The variation of the gap with $\Lambda_b=L/2 \pi^2 \xi_b$ is shown on
Fig.({\ref{fig3}). $\xi_b$ is the coherence length of the infinite
system. When the number of particles
is even, one sees that the gap {\it increases} with decreasing size.
 This seems to be in contradiction with ref.\cite{Visscher96} who found
a decrease of the critical temperature $T_c$ when the size decreases. We
will comment on this point
in the next section. When the number of particles is odd, the gap
decreases with decreasing size.
 When the length becomes too small, the
order parameter can even vanish. This happens for a
critical value of $\Lambda_b= 1 / 2 \gamma= .280$. $\gamma$ is the
Euler constant $\gamma=1.781$.

\begin{figure}[hbt]
\centerline{
\epsfxsize 8cm
\epsffile{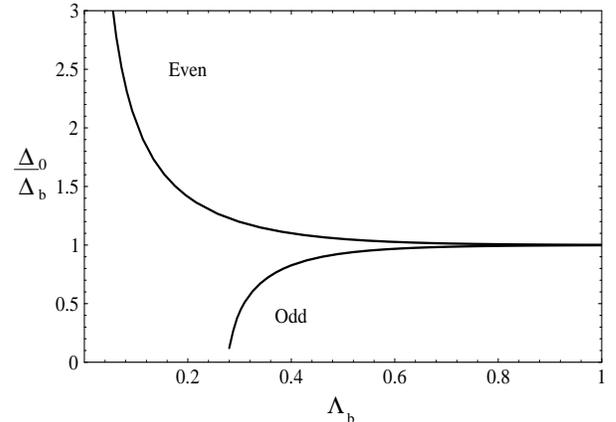}
}
\caption{Variation of the dimensionless gap $\Delta_0 / \Delta_b$ with
$\Lambda_b=\Delta_b/2 \delta$. $\Delta_b$ is the gap of the infinite
system. $\Lambda_b$ is proportional to the size $L$.}
\label{fig3}
\end{figure}

We now turn  to the effect of the AB flux.
In a finite flux, the gap equation (\ref{delta0}) transforms into:
\begin{equation}
\ln {\Delta_\varphi \over \Delta_b} = 2 \sum_{m>0}
(-1)^{Nm} K_0(2 \pi m {\Delta_\varphi \over \Delta_b} \Lambda_b)  \cos 2 \pi
m \varphi
\label{deltaphi}
\end{equation}
Fig.(\ref{fig4}) shows the gap in the case of even $N$, for different
fluxes. It is seen
 the effect of the flux is to reduce the gap, this effect being larger for
small sizes. Actually the case $\varphi=1/2$ with an even number of
particles is equivalent to the case $\varphi=0$ with an odd number of
particles. This is obvious in Figs.(\ref{fig3}) and (\ref{fig4}) and in
the structure of the gap equation (\ref{gap3}).

\begin{figure}[hbt]
\centerline{
\epsfxsize 8cm
\epsffile{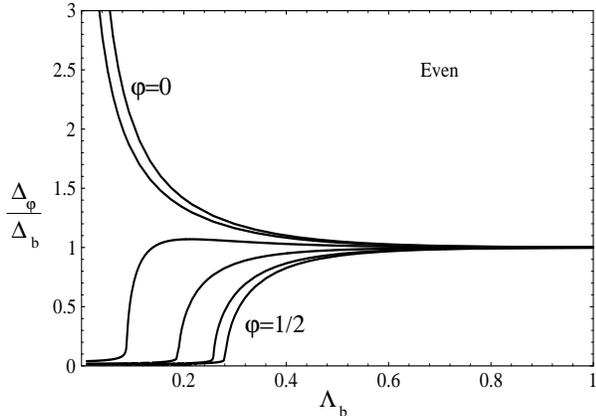}
}
\caption{Variation of the dimensionless gap $\Delta_\varphi / \Delta_b$ with
$\Lambda_b$, for different fluxes $\varphi=0,0.1,0.2,0.3,0.4,0.5\ .$}
\label{fig4}
\end{figure}

Fig.(\ref{fig5}) shows the flux dependence of $\Delta_\varphi /
\Delta_b$
with the flux for different parameters $\Delta_b/ 2 \delta \propto L /
\xi_b$. For an even number of
particles, the order parameter is {\it enhanced} at zero flux and is
reduced for large flux. When $\Delta_b / 2 \delta$ becomes too small, the
order parameter can even vanish near $\varphi=1/2$. This happens for a
critical value of $\Lambda_b= 1 / 2 \gamma= .280$. For smaller
rings, i.e. when $\Lambda_b$ is smaller, the CDW disappears at a critical
flux given by:$$\psi(\varphi_c)+\psi(1-\varphi_c)=2 \ln \Lambda_b/2$$where
$\psi$ is the digamma function. \begin{figure}[hbt]
\centerline{
\epsfxsize 8cm
\epsffile{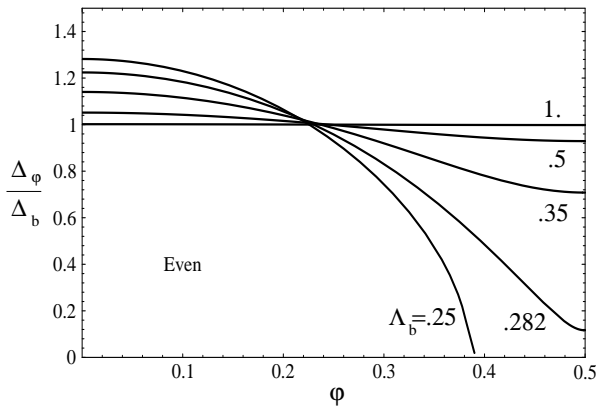}}
\centerline{
\epsfxsize 8cm
\epsffile{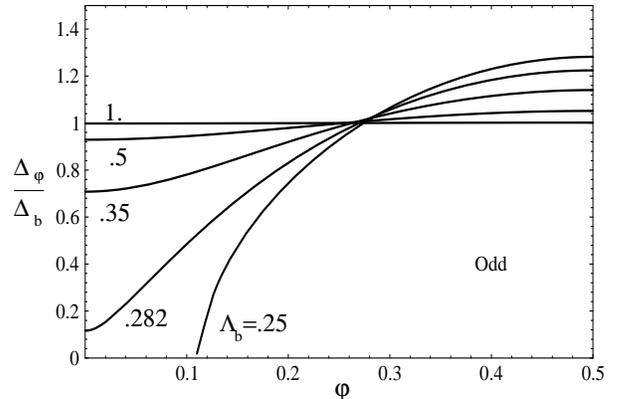}}
\caption{Dimensionless gap $\Delta_\varphi / \Delta_b$ versus flux for
different system sizes a) with even $N$, b) with odd $N$.   }
\label{fig5}
\end{figure}

For the case of even N, the figure (\ref{fig5}.a) is consistent with the
figure 2 of ref.\cite{Visscher96} where $\Delta_\varphi / \Delta_0$ was
plotted instead of $\Delta_\varphi / \Delta_b$ here. That figure could not
show the interesting result that in low flux the gap
increases when the size decreases.

For odd $N$, the order parameter {\it increases}
 in a finite magnetic flux and it is maximum at $\varphi=1/2$.
This contradicts the arguments of ref.\cite{Visscher96} who argues
 that the suppression of the CDW order is due to pair-breaking induced
 by the field. {\it There is indeed no pair-breaking effect since the field
 does not couple to the phase of the electron-hole pair}. The field
effect here is simply to change the position of the energy levels
and thus to  {\it either reduce or enhance} the stability of the CDW.
This can be simply understood from the schematic Figs.(\ref{fig2}).

 For a ring of large size, the modulation of the gap given by eq.\ref{deltaphi} becomes weak and harmonic.
% $K_0(x) =\sqrt{\pi /2 x}e^{-x}$ for large $x$,
 It is given by:
$$\Delta_\varphi =\Delta_b ( 1 \pm {1 \over \sqrt{\Lambda_0}} e^{-2
\pi \Lambda_0}  \cos 2 \pi \varphi)$$
which explicitely displays the exponential decrease of the  modulation with
the size of the ring.
\section{Transition temperature} \label{TRAN}
The dependence of the critical temperature with the size and the flux
reflects those of the gap.

At the transition, $\Delta = \Lambda =0$. As in ref.\cite{Visscher96}, the
self-consistency equation
can  be written as, for even $N$:

$$
{1 \over g} ={1 \over 2} \sum_n {
\tanh[{{\beta'_\varphi \over 2} }(n+\varphi)] \over n+\varphi
} $$

Doing the same Poisson summation as above, the self consistency equation for
the critical temperature $T_\varphi$ is found to be, taking into account
the parity:
 \begin{equation}{1 \over g} = \ln 1.14 {E^* \over T_\varphi}+\sum_{m>0}
(-1)^{Nm} F(m/\beta'_\varphi) \cos(2 \pi m \varphi)\label{T}\end{equation}
where the function $F$ is:
$$F(x)=  \ln \left( {\cosh (2 \pi^2 x)
  +1 \over \cosh (2 \pi^2 x) -1}\right) $$

The critical temperature of the infinite system is given by:

\begin{equation}{1 \over g} = \ln 1.14 {E^* \over
T_b}\label{Tinf}\end{equation}
so that by difference between eqs.(\ref{T}) and (\ref{Tinf}), one has the
flux
dependence of $T_\varphi/T_b$ for different parameters $\Lambda_b= \Delta_b
/ 2 \delta$:
\begin{equation}\ln {T_\varphi \over T_b}  =  \sum_{m>0}
(-1)^{Nm} F\left({m\over 1.76} {\Lambda_b } {T_\varphi \over
T_b} \right) \cos 2 \pi m \varphi
\label{TTinf}\end{equation}
The result is shown on Fig.(\ref{fig6}). For zero flux, the variation is
very similar to
that of the gap (Fig.\ref{fig3}). When the
size decreases, the critical temperature increases, in apparent contradiction
with the figure (4) of ref.\cite{Visscher96} who found a decrease of the
critical temperature.
However their temperature is normalized to the gap, $\Delta_0$ which is
itself size
dependent, and not to $\Delta_b$. What is actually found in
ref.\cite{Visscher96} is an increase
of the BCS ratio $\Delta_\varphi /T_\varphi$. This is in agreement with our
calculation of this
ratio shown in Fig.(\ref{fig7}) and exhibits the $12\%$ increase found in
ref.\cite{Visscher96}.
\begin{figure}[hbt]
\centerline{ \epsfxsize 8cm
\epsffile{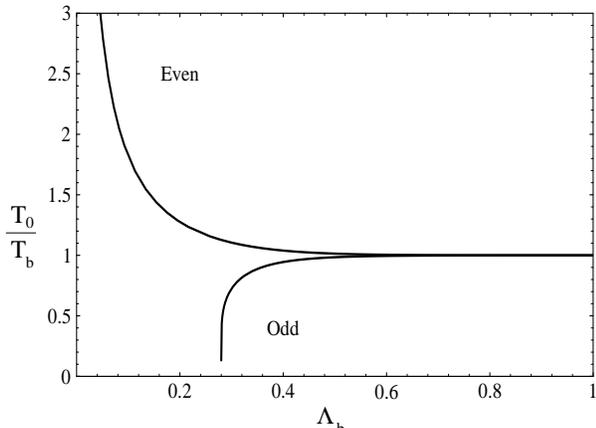}
}
\caption{Variation of the critical temperature with the size, in zero flux.}
\label{fig6}
\end{figure}

 \begin{figure}[hbt]
\centerline{
\epsfxsize 8cm
\epsffile{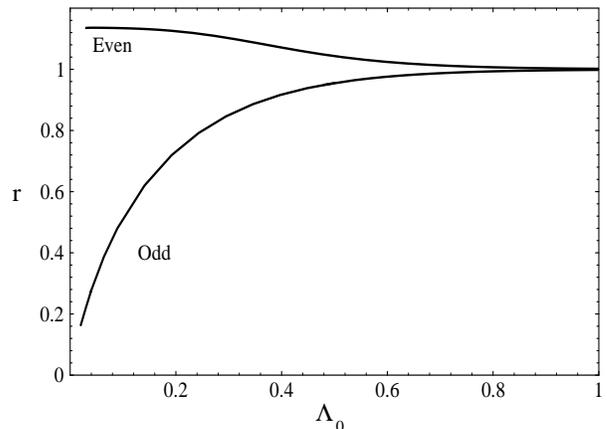}
}
\caption{Variation of the normalized BCS
 ratio $r=(\Delta_0 /T_0)/ (\Delta_b /T_b)$ with
 the size, in zero flux}
\label{fig7}
\end{figure}

Finally, we plot in Fig.(\ref{fig8}) the variation of the critical
temperature with the flux, for different sizes.

\begin{figure}[hbt]
\centerline{ \epsfxsize 8cm
\epsffile{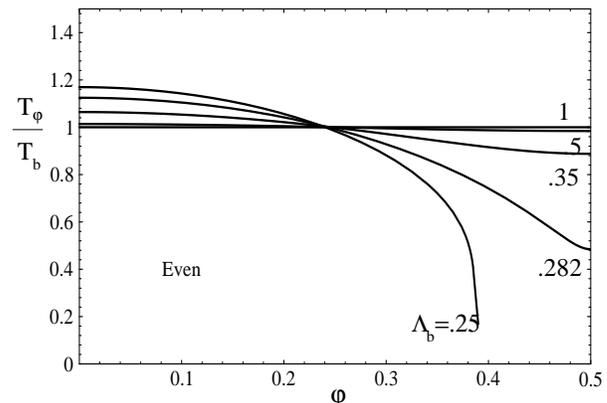}
}
\caption{Variation of the dimensionless critical temperature
$T_\varphi/T_b$ with the flux, for different system sizes, when $N$
is even. The case where $N$ is odd deduces by the same symmetry as in Fig. 5
}
\label{fig8}
\end{figure}

In the limit of a large system, the oscillations of the critical
temperature become exponentially small as:

$$T_\varphi = T_b ( 1 \pm 2 e^{-1.14 L / \xi} \cos 2 \pi \varphi) $$

\section{Persistent current}\label{PER}

The persistent current in the CDW phase is given by:
\begin{equation}I(\varphi) = -\sum_ n 2 {I_0 \over N} (n+ \varphi) + \sum_n
I_0 {n+\varphi \over \sqrt{(n+\varphi)^2 +
\Lambda^2}}\label{current}\end{equation}
where $n \in [ -N/2,N/2-1]$ if $N$ is even and $n \in [
-(N-1)/2,(N-1)/2]$ if $N$ is odd.
$$I_0= {2 \delta \over \phi_0} = {e v_F \over L}$$ is the maximal
current in one dimension.  Eq.(\ref{current}) for the
persistent
current is exact for a quadratic dispersion relation\cite{quad}. The first
term is the
persistent current $I_N(\varphi)$ in the normal state: $I_N(\varphi)=-2 I_0
\ \varphi$ when $N$ is odd and
  $I_N(\varphi)=-I_0 \ (2 |\varphi| -1)$  when $N$ is even. \cite{comment}

 After
summation
by parts and  Poisson summation, the flux dependence of the total current
can be conveniently cast in the Fourier expansion which is parity dependent:
$$I(\varphi) = 4 I_0 \Lambda_\varphi \sum_{m>0} (-1)^{Nm} K_1(2 \pi m
\Lambda_\varphi) \sin 2 \pi m  \varphi $$
$K_1$ is a modified Bessel function of the second
kind\cite{Bessel}.
The persistent current depends on the gap which is itself flux
dependent.  For pedagogical purpose, we first show the flux
variation of the current {\it assuming a constant gap},
Fig.(\ref{fig9}).
\begin{figure}[hbt]
\centerline{ \epsfxsize 8cm
\epsffile{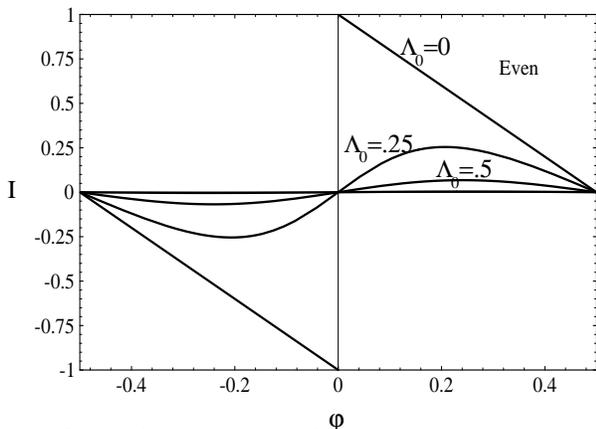}
}
\caption{Persistent current with a constant gap $\Delta_0$, for an even
number of particles.}
\label{fig9}
\end{figure}
When the gap goes to zero,
 $K_1(x) \rightarrow 1/x$ and one
recovers the current of the normal state\cite{Cheung89}:

$$I(\varphi) =  {2 \over \pi} I_0 \sum_{m>0} {(-1)^{Nm} \over m}
 \sin 2 \pi m  \varphi $$
When the gap becomes larger than the interlevel spacing,
the current is  reduced  exponentially as:

$$I(\varphi) =(-1)^{N}  I_0   e^{-1.14 L / \xi} \sin 2 \pi \varphi $$

The variation of the gap itself with the flux must be taken into account.
Fig.(\ref{fig10}a) shows the variation of the current with the flux for
$\Lambda_0=.35$. When  the flux increases, the gap decreases  so that the
current becomes larger (full line) than if the gap were constant (dashed
line). When $\Lambda_0=.25$, the CDW gap vanishes at a critical flux and
the current recovers {\it continuously} its value in the metallic phase. This
results constradicts those of ref. \cite{Yi97} who found a discontinuity in
the current. When $N$ is odd, the current is trivially shifted by half a
period $\phi_0/2$, at variance with the conclusion of ref.\cite{Yi97}.
\begin{figure}[hbt]
\centerline{ \epsfxsize 8cm \epsffile{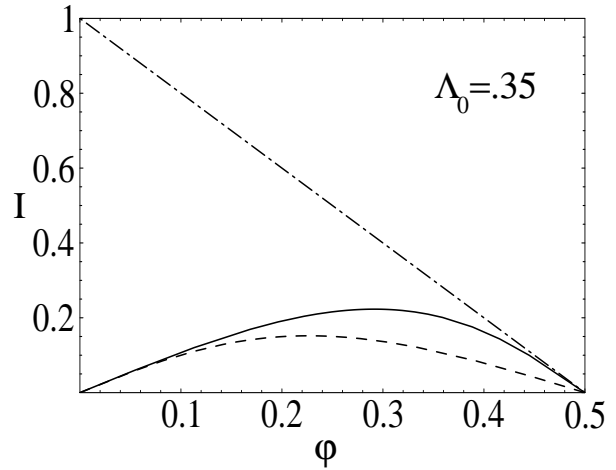}}
\centerline{ \epsfxsize 8cm \epsffile{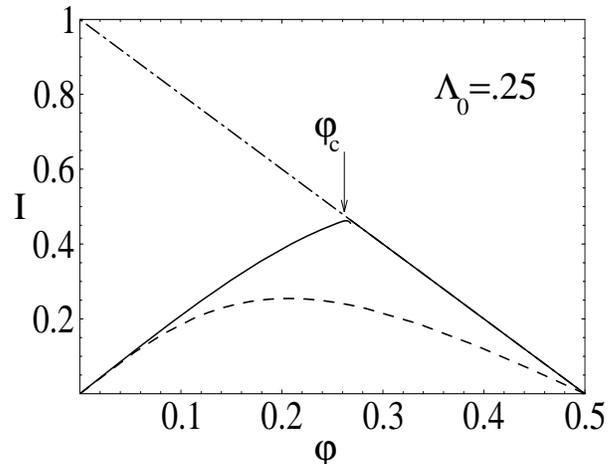}}
\caption{Persistent current in the CDW phase (full line), for two values of
the size $L$, in the case of even $N$. The dashed line shows the current if
the gap were constant and the dotted-dashed line shows the current in the
metallic phase.}
\label{fig10}
\end{figure}
\section{Conclusions}    \label{CON}

We have derived the mean-field thermodynamics of a CDW in a small 1D clean
system in the presence of a magnetic flux.
The stability of the CDW  depends on the parity of the number $N$ of
particles. When the size decreases and becomes of the order of the
coherence length $\xi=\hbar v_F / \pi \Delta$,
the CDW order parameter increases if $N$ is even, it decreases if $N$ is odd.
	   The CDW is stabilized by the magnetic flux when $N$ is odd and it
is destabilized when $N$ is even.
These results correct those of refs.\cite{Visscher96,Yi97} who found that
the flux always tends to suppress the Peierls instability.

These are  the results for a one dimensional ring.
They can be in principle generalized to
the case of a many channel ring. Ref.\cite{Visscher96} suggests that the
current is simply multiplied by the number of chains. This is not true, as
it is already known for the metallic phase that the current
results from a subtle addition of the contributions of the
different channels\cite{Cheung88}. Such rings with few number of channels
can be synthetized using thin-film growth of blue bronze
oxydes\cite{Zant96,Mantel97}.
The case of a 1D ring with short range interaction and impurities, a
disordered Luttinger liquid, has been studied recently\cite{Giamarchi95}. It
would be interesting to see how the discretness of the spectrum affects the
obtained results.

Note added in proof: After this paper was accepted, I have been informed
by F. Von Oppen of the existence of a related work with similar
conclusions\cite{Nathanson92}. Here I have found
the analytical expressions of the harmonics expansion of the critical
temperature, gap and persistent current. They agree with previous numerical
calculations. Ref.\cite{Nathanson92} addresses the fluctuation effect.


\begin{thebibliography}{99}
\bibitem{quad} A linear dispersion relation for the normal
phase would give a zero current in the normal phase.
\end{thebibliography}
 \end{document}